# Casimir vacuum energy and the semiclassical electron

H. E. Puthoff

Institute for Advanced Studies at Austin, 11855 Research Blvd., Austin, TX 78759, USA

E-mail: Puthoff@earthtech.org

**Abstract**

In 1953 Casimir proposed a semiclassical model for the electron based on the concept that net inward radiation pressure from the electromagnetic vacuum fluctuations fields (as in the Casimir effect, generally) might play the role of Poincare stresses, compensating outward coulomb pressure to yield a stable configuration at small dimensions. Given that in scattering experiments the electron appears point-like, critical to the success of the proposed model is demonstration that the self-energy corresponding to the divergent coulomb field does not contribute to the electron mass. Here we develop a self-consistent, vacuum-fluctuation-based model that satisfies this requirement and thereby resolves the issue of what would otherwise appear to be an incompatibility between a point-like electron and finite mass.

## 1. Introduction

In his paper "Introductory Remarks on Quantum Electrodynamics" [1], H.B.G. Casimir, fresh from his investigation of the vacuum-fluctuation-driven attractive force between conducting plates (the Casimir force), proposed a semiclassical model of the electron in the spirit of Lorentz's theory of the electron [2]. Though such modeling is out of favor in terms of providing a realistic description of the quantum electron, nonetheless the potential for insights to emerge from such modeling remains seductive. Along these lines Casimir suggested two models, the second of which has received little attention but is the subject of this short note.

Casimir suggested that a dense shell-like distribution of charge might partially (as in the Casimir Effect, generally), or even wholly, suppress vacuum fields in the interior of the shell (Shell Models I and II, respectively). This could result in net inward radiation pressure from the EM (electromagnetic) vacuum fluctuation fields (playing the role of Poincare stresses) compensating outwardly-directed coulomb forces to yield a stable configuration at small dimensions. From the modern viewpoint, for such semiclassical modeling to provide useful insight, a key issue to be resolved is that although in scattering experiments the electron appears to be pointlike and structureless, the self-energy corresponding to the (divergent) coulomb field does not appear to contribute to the electron mass. In QED renormalization theory this is handled by adding an infinite negative mass term to compensate the infinite positive coulomb term. Though such renormalization can be carried out in QED in an unambiguous and invariant way, from the standpoint of a semiclassical model it appears sufficiently *ad hoc* as to merit a search for an alternative.

**2. Casimir shell model I**

In Casimir's Shell Model I, interior EM vacuum fluctuation fields associated with the discrete vacuum states permitted by boundary conditions are assumed to exist (as in the standard modeling of the Casimir Effect). For this model a charge *e* is assumed to be homogeneously distributed on the surface of a conducting shell of vanishing radius *a*, whose tendency to expand by coulomb repulsion is checked by a net inwardly-directed vacuum fluctuation radiation pressure due to the pressure imbalance between interior and exterior vacuum modes. Unfortunately for the model, detailed analyses showed that for this case the Casimir pressure was (surprisingly) outwardly-directed, augmenting rather than offsetting the coulomb pressure, thereby falsifying the Casimir Poincare stress hypothesis for this case [3-6].

**3. Casimir shell model II**

For Casimir's Shell Model II, which we now discuss, the charge density is taken to be sufficiently dense in a vanishing-radius shell so as to result in the total absence of interior vacuum fluctuation fields as a singularity is approached. The modeling for this case proceeds as follows:

1. Consider charge *e* to be homogeneously distributed on a spherical shell of (vanishing) radius $a \to 0$.

2. Under this assumption, the electric field *E*, given by

$$E = \left(\frac{e}{4\pi\varepsilon_0 r^2}\right), \tag{1}$$

leads to a (formally-divergent) coulomb energy

$$W_{coul} = \lim_{a \to 0} \int_a^\infty u_{coul} dV = \lim_{a \to 0} \int_a^\infty \frac{1}{2}\varepsilon_0 E^2 dV = \lim_{a \to 0}\left(\frac{e^2}{8\pi\varepsilon_0 a}\right) = \lim_{a \to 0}\left(\frac{\alpha\hbar c}{2a}\right), \tag{2}$$

where $\alpha$ is the fine structure constant, $\alpha = e^2/4\pi\varepsilon_0 \hbar c \approx 1/137.036$.

3. With regard to the vacuum fluctuation electromagnetic fields, the spectral energy density given by

$$\rho(\omega)d\omega = \frac{\hbar\omega^3}{2\pi^2 c^3}d\omega \qquad (3)$$

leads to an associated divergent energy density

$$u_{vac} = \lim_{\Omega\to\infty}\left(\int_0^{\Omega}\frac{\hbar\omega^3}{2\pi^2 c^3}d\omega\right) = \lim_{\Omega\to\infty}\left(\frac{\hbar\Omega^4}{8\pi^2 c^3}\right), \qquad (4)$$

where $\Omega$ constitutes an upper limit cutoff frequency that asymptotically approaches infinity. For Shell Model II in which an absence of interior vacuum fluctuation energy is assumed, the vacuum energy deficit inside the sphere is given by

$$W_{vac} = -u_{vac}V = \lim_{\Omega\to\infty, a\to 0}\left(-\frac{\hbar\Omega^4}{8\pi^2 c^3}\cdot\frac{4}{3}\pi a^3\right) = \lim_{\Omega\to\infty, a\to 0}\left(-\frac{\hbar\Omega^4 a^3}{6\pi c^3}\right). \qquad (5)$$

From (2) and (5) we therefore obtain for the coulomb and vacuum energy contributions to the shell model

$$W = W_{coul} + W_{vac} = \lim_{a\to 0}\left(\frac{\alpha\hbar c}{2a}\right) - \lim_{\Omega\to\infty, a\to 0}\left(\frac{\hbar\Omega^4 a^3}{6\pi c^3}\right). \qquad (6)$$

4. We now require that the outwardly-directed coulomb pressure, given by

$$P_{coul} = u_{coul} = \lim_{a\to 0}\left(\frac{\alpha\hbar c}{8\pi a^4}\right), \qquad (7)$$

be balanced by the inwardly-directed vacuum radiation pressure

$$P_{vac} = -\frac{1}{3}u_{vac} = \lim_{\Omega\to\infty}\left(-\frac{\hbar\Omega^4}{24\pi^2 c^3}\right). \qquad (8)$$

Pressures (7) and (8) result in a (stable) balance at radius $a = a_b$ given by

$$a_b = \lim_{\Omega\to\infty}\left[\frac{(3\pi\alpha)^{1/4} c}{\Omega}\right] \to 0. \qquad (9)$$

5. Under this pressure-balance condition, the shell model energy (6) then reduces to

$$W = W_{coul} + W_{vac} = \lim_{a_b\to 0}\left(\frac{\alpha\hbar c}{2a_b}\right) - \lim_{a_b\to 0}\left(\frac{\alpha\hbar c}{2a_b}\right) \equiv 0. \qquad (10)$$

## 4. Discussion and conclusions

Thus, for Casimir's Shell Model II, the net contribution to the self-energy of the point-particle electron by the coulomb and vacuum fields vanishes. We are therefore led to conclude that, under the set of assumptions applicable to Casimir's Shell Model II, an inwardly-directed, divergent, electromagnetic vacuum fluctuation radiation pressure stably balances the divergent coulomb pressure. Furthermore, it does so in such a manner that, even in the limiting case of the point particle electron, no contribution to the self energy of the electron results from the divergent coulomb field. Thus a key requirement for the semiclassical electron model is met. As a result, to the degree that this result of the semiclassical analysis carries over to QED renormalization, it would appear that the additive infinite negative mass in the QED approach finds its source in a negative vacuum energy contribution as proposed in the Casimir model. Finally, the reality of high-energy-density vacuum fluctuation fields at the fundamental particle level is buttressed, while at the same time leading to a renormalization process compatible with a finite particle mass.

**Acknowledgment**

I wish to express my appreciation to Michael Ibison for stimulating discussion and useful input during the development of this effort.